\documentclass[3p,times,procedia]{elsarticle}
\usepackage{nupha_ecrc}
\usepackage{multirow}

\usepackage{amsmath}
\usepackage{amssymb}
\usepackage{multirow}
\usepackage{graphicx}
\usepackage{subfigure}
\usepackage{bbm}

\volume{00}

\firstpage{1}

\journalname{Nuclear Physics A}

\runauth{}

\jid{nupha}

\jnltitlelogo{Nuclear Physics A}

\usepackage{amssymb}

\usepackage{lineno}
\usepackage{wrapfig}                                                            
\usepackage{placeins}
\usepackage{multicol}
\newcommand{\dd}{\mathrm{d}}

\usepackage[figuresright]{rotating}

\begin{document}

\begin{frontmatter}



\dochead{}

\title{Charmonium and bottomonium spectral functions in the vector channel}


\author[label1]{H.-T.~Ding}
\author[label1,label2]{O.~Kaczmarek}
\author[label2]{A.-L.~Kruse}
\author[label3]{R.~Larsen}
\author[label2]{L.~Mazur} 
\author[label3]{Swagato~Mukherjee}
\author[label4]{H.~Ohno}
\author[label2]{H.~Sandmeyer}

\author[label1,label2]{H.-T.~Shu\corref{cor1}$^*$}

\address[label1]{Key Laboratory of Quark and Lepton Physics (MOE) and Institute of
    Particle Physics, \\ Central China Normal University, Wuhan 430079, China}
\address[label2]{Fakult\"at f\"ur Physik, Universit\"at Bielefeld, 33615 Bielefeld, Germany}
\address[label3]{Physics Department, Brookhaven National Laboratory, Upton, NY 11973, USA}
\address[label4]{Center for Computational Sciences, University of Tsukuba, Tsukuba, Ibaraki 305-8577, Japan}

\cortext[cor1]{Speaker}

\begin{abstract}

In this paper we report our results on quarkonium spectral functions in the vector channel obtained from quenched lattice QCD simulations at $T\in[0.75, 2.25]~T_c$. The calculations have been performed on very large and fine isotropic lattices where both charm and bottom quarks can be treated relativistically. The spectral functions are reconstructed using the Maximum Entropy Method. We study the dissociation of quarkonium states from the temperature dependence of the spectral functions and estimate heavy quark diffusion coefficients using the low-frequency behavior of the vector spectral functions.

\end{abstract}


\end{frontmatter}


\vspace{1cm}

In the past decade many efforts have been made to gain a deep insight into the properties of the hot medium produced in heavy ion collisions~\cite{Brambilla:2014jmp}. Heavy quarkonium, which are produced in the early stage of the collision and involved in  the whole evolution of the system, are ideal probes in studying the hot medium due to their large mass~\cite{Andronic:2015wma} compared with the temperature of the surrounding plasma. In a recent run at the CMS collider an obvious sequential suppression of the excited $\Upsilon$ states has been observed in the $Pb+Pb$ collisions~\cite{Sirunyan:2017lzi}. There are several works trying to figure out how this happens from different perspectives e.g. non-relativistic QCD~\cite{Aarts:2016nwb,Kim:2014iga}, potential non-relativistic QCD~\cite{Brambilla:2010vq}, T-matrix approach~\cite{Liu:2017qah} and perturbative calculations~\cite{Burnier:2017bod} etc., however the results still remain inconclusive. In principle the fate of the hadrons in the medium can be read off directly from the deformation of the resonance peaks in the hadron spectral functions. However a so-called ``transport peak'' appearing in the small frequencies of the spectral function in the vector channel makes the reconstruction of the resonance peaks more complicated. This transport peak is directly related to the heavy quark diffusion coefficient which is a crucial input in hydro and transport models to describe the experimental data. The heavy quark diffusion coefficient has been calculated perturbatively. However, in the interesting temperature range the results from leading order and next leading order calculations differ a lot~\cite{Moore:2004tg,CaronHuot:2007gq}. 

To understand the temperature dependence of spectral functions from first principle, we carry out lattice QCD simulations and report the results in this proceedings. To start we calculate the Euclidean 2-point ``current-current'' temporal correlation functions in the vector channel on the lattice. They are related to the hadron spectral functions (SPF) $\rho_{\mu}$ as follows
\begin{align}
    \label{eqn_integ_trans}
    G_{\mu}(\tau,\vec{p},T)\equiv \sum_{x,y,z}e^{(-i\vec{p}\cdot\vec{x})}\ \langle J_{\mu}(0,\vec{0})\ J^{\dagger}_{\mu}(\tau,\vec{x})\rangle=\int\limits_0^{\infty}
    \frac{\dd \omega}{2\pi}\ \rho_{\mu}(\omega,\vec{p},T)\ K(\omega,\tau,T),
\end{align}
where $K(\omega,\tau,T)=\cosh(\omega\tau-\omega/2T)/\sinh(\omega/2T)$ is the integrand kernel. The heavy quark diffusion coefficient can be obtained from the vector spectral function through the relation $D=\frac{1}{6\chi_{00}}\lim_{\omega\rightarrow 0}\sum_{\mu=1}^{3}\frac{\rho_{\mu}(\omega,\vec{0})}{\omega}$, where $\chi_{00}$ is the quark number susceptibility~\cite{Petreczky:2005nh}. To extract the continuous spectral function from the discretized correlation functions we adopt the Bayesian theorem based Maximum Entropy Method (MEM)~\cite{Asakawa:2000tr}.


We apply MEM to correlation functions obtained on large quenched isotropic lattices at fixed spatial extent $N_{\sigma}=192$ and five different temporal extents $N_{\tau}=96, 64, 56, 48$ and $32$, which corresponds to temperature $T=0.75, 1.10, 1.25, 1.50$ and $2.25T_c$, respectively. The gauge coupling $\beta$ is 7.793 and the corresponding lattice spacing is $a=0.009$ fm~\cite{Sommer:1993ce}. We have performed simulations in a wide range of $\kappa$ values~\cite{Ding:2017rty}, of which the largest one 0.13221 and smallest one 0.12798 can reproduce nearly physical $J/\psi$ mass and $\Upsilon$ mass (the resultant screening mass are 3.38 GeV and 10.47 GeV for $J/\psi$ and $\Upsilon$, respectively) and we only present these two in this work. In our simulations we use the Wilson gauge action and the non-perturbatively Clover-improved Wilson fermions~\cite{Sheikholeslami:1985ij}. The relative errors of the obtained correlation functions $\delta G/G$ at the middle point are $\sim0.3$\%.




Due to the general problem that the number of data points is not sufficient in the MEM analysis we are always careful about the dependence of the output spectral functions on the default models and the number of data points in the correlation functions. In our work we examine the influence of these two factors carefully one by one. In the large frequency part we use the free lattice spectral function for Wilson fermions~\cite{Karsch:2003wy,Aarts:2005hg} to accomodate the lattice cutoff effects. The transport peak we parametrize as a Breit-Wigner distribution according to kinetic theory~\cite{Petreczky:2005nh}. The resonance peak is parametrized as a relativistic Breit-Wigner distribution. To get rid of the trivial temperature dependence of the integrand kernel $K(\omega,\tau,T)$, we analyze the reconstructed correlation function $G_{rec}(\tilde\tau,T;T')$ defined as a correlation function reconstructed \setlength{\columnsep}{4pt}
\begin{wrapfigure}[19]{r}{0.66\textwidth}\vspace{-0.6cm}
\begin{center}
     \includegraphics[width=0.33\textwidth]{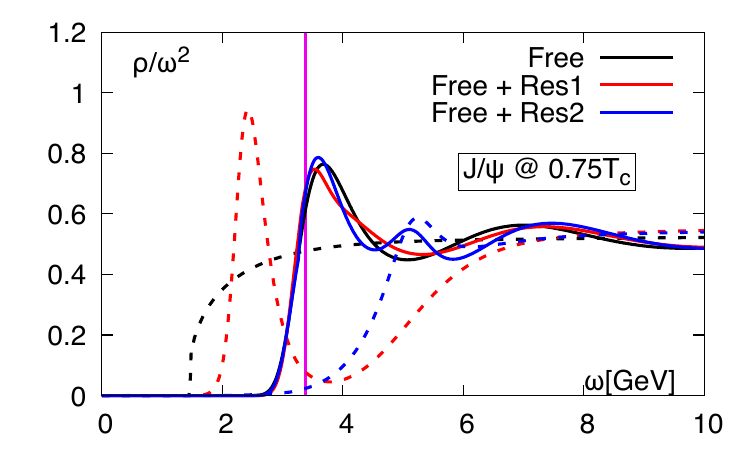}\includegraphics[width=0.33\textwidth]{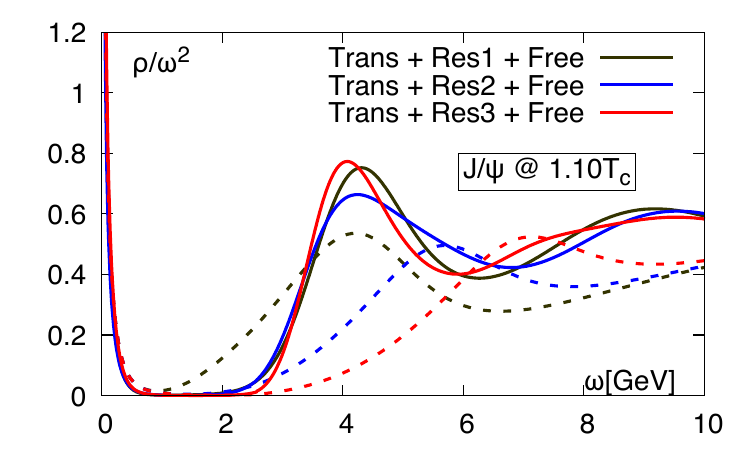}
     \includegraphics[width=0.33\textwidth]{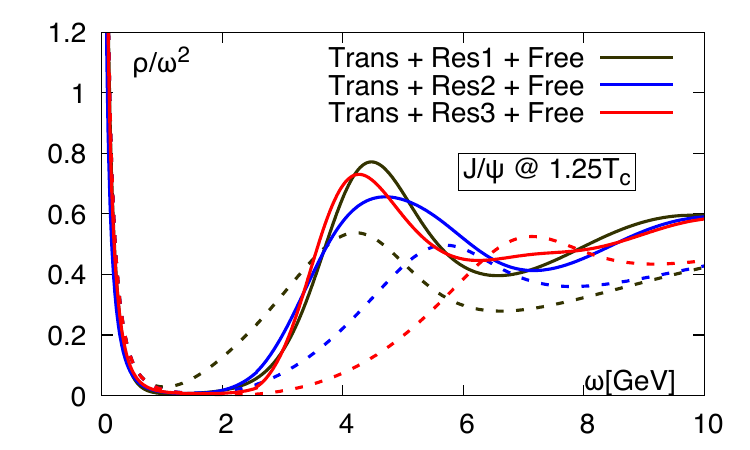}\includegraphics[width=0.33\textwidth]{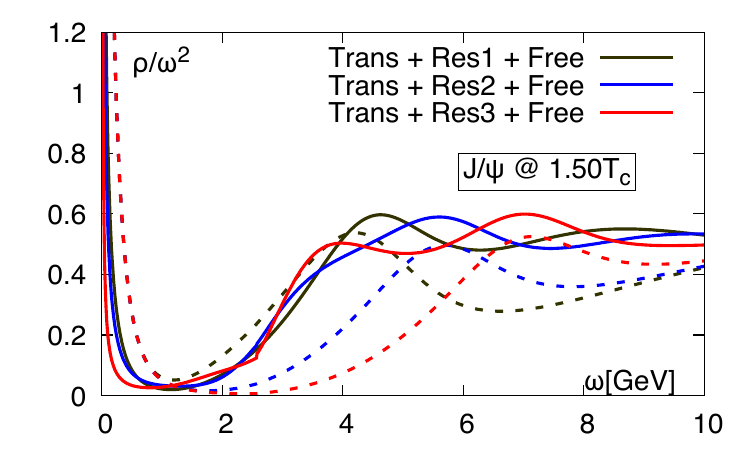}
  \end{center}\vspace{-0.5cm}
 \caption{Charmonia spectral functions at different temperatures obtained by MEM with different default models. \textit{Top left}: $0.75T_c$. \textit{Top right}: $1.10T_c$. \textit{Bottom left}: $1.25T_c$. \textit{Bottom right}: $1.50T_c$. $T_c$ is the confinement/deconfinement transition temperature. The vertical line indicates the screening mass of $J/\psi$.}\label{charm_dm}
 \end{wrapfigure} from the spectral function obtained already at a lower temperature $T'$ and the integrand kernel at another $T$~\cite{Ding:2017std}. The reconstructed correlation function $G(\tau,T;T')$ and the original correlation function $G(\tau,T)$ have the same number of data points and only differ in the spectral function. Thus one can compare the spectral functions extracted from these two correlation functions and study whether the change of the spectral function from one temperature to another suffers from the reduction of data points in the correlation functions.

In the following we will show the results of spectral functions obtained using MEM. In all figures of the spectral functions in this paper, the dashed curves are the default models and the solid curves with the same color are corresponding output spectral functions. 

 \setlength{\columnsep}{4pt}
\begin{wrapfigure}[12]{r}{0.66\textwidth}\vspace{-0.8cm}
\begin{center}
     \includegraphics[width=0.33\textwidth,height=3.5cm]{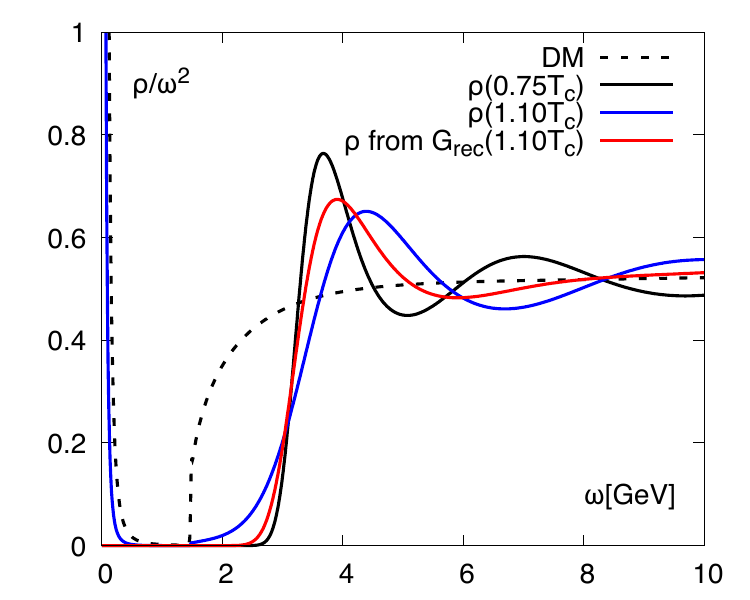}\includegraphics[width=0.33\textwidth,height=3.5cm]{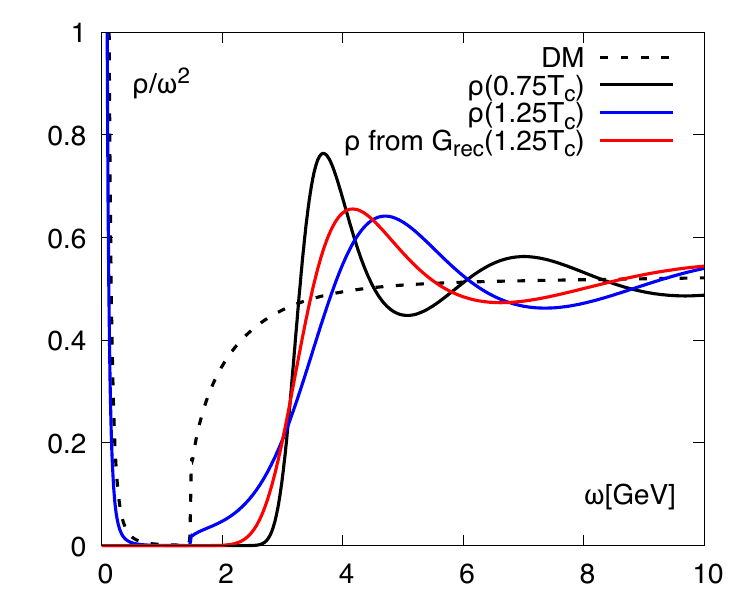}
  \end{center}\vspace{-0.5cm}
 \caption{Comparison of charmonia spectral functions obtained from the original correlation functions and the reconstructed correlation functions. \textit{Left}: $T=1.10T_c, T'=0.75T_c$. \textit{Right}: $T=1.25T_c, T'=0.75T_c$.}
\label{charm_rec}
\end{wrapfigure}

First we show the charmonia spectral functions in Fig.\ref{charm_dm}. At 0.75$T_c$, as shown in the left top panel, the spectral function is believed to have no transport peak, therefore the first default model is just a free lattice spectral function while the second and third are of the same type: a combination of a free lattice spectral function and a resonance peak connected by a modified $\Theta$-function~\cite{Ding:2017std,Ding:2016hua}. The difference is that their resonance peak locations are respectively smaller and larger than the screening mass of $J/\psi$ at $0.75T_c$. We found that although the default models are quite different, the first low-lying peak is stable and its location is close to the screening mass of $J/\psi$ (the magenta vertical line in the figure). 

\setlength{\columnsep}{4pt}
\begin{wrapfigure}[13]{r}{0.36\textwidth}\vspace{-0.6cm}
  \begin{center}
     \includegraphics[width=0.36\textwidth,height=3.9cm]{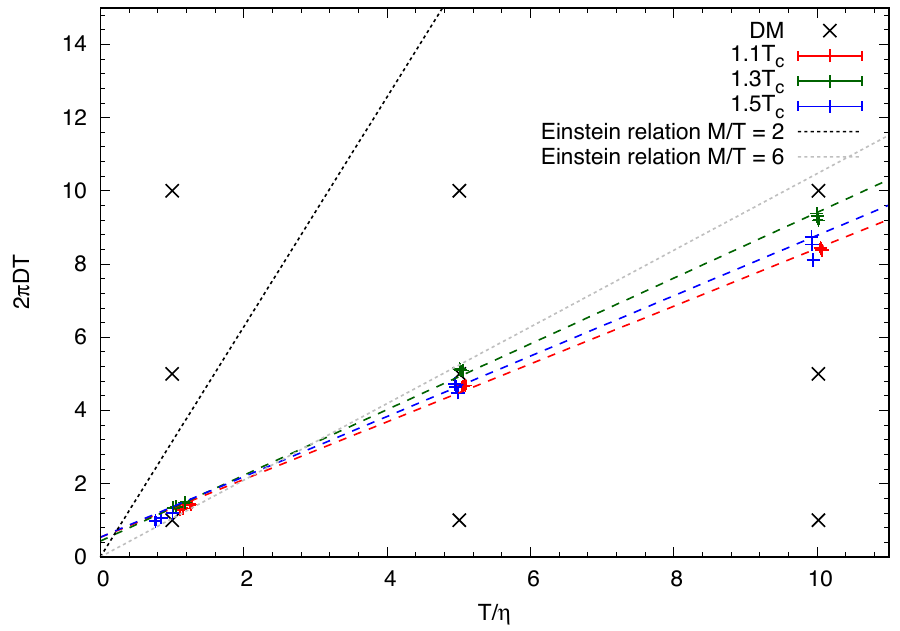}                                                                                                    
  \end{center}\vspace{-0.5cm}
 \caption{Linear relation between $2\pi TD$ and $T/\eta$ at 1.10, 1.25 and 1.50$T_c$ for charmonia.}\label{2piTD}
 \end{wrapfigure} 

At 1.10$T_c$ shown in the top right panel the three default models are of the same type: a combination of a transport peak, a resonance peak and a free lattice spectral function. The difference from the default models at $T<T_c$ is that they have larger and larger resonance peak locations. We found that the first peak is still stable but moves to a higher frequency region. At 1.25$T_c$ and 1.50$T_c$ the default models used are exactly the same as the ones used at $1.10T_c$. As shown in the bottom panel the output spectral functions at 1.25$T_c$ are very similar as those obtained at $1.10T_c$ but the peak locations are shifted to even higher frequencies while at 1.50$T_c$ the spectral functions become flat. The reason of the observed strong default model dependence could be the lack of statistics or data points in the correlation functions which needs further investigations.

\setlength{\columnsep}{4pt}
\begin{wrapfigure}[20]{r}{0.66\textwidth}\vspace{-0.8cm}
\begin{center}
     \includegraphics[width=0.33\textwidth,height=3.5cm]{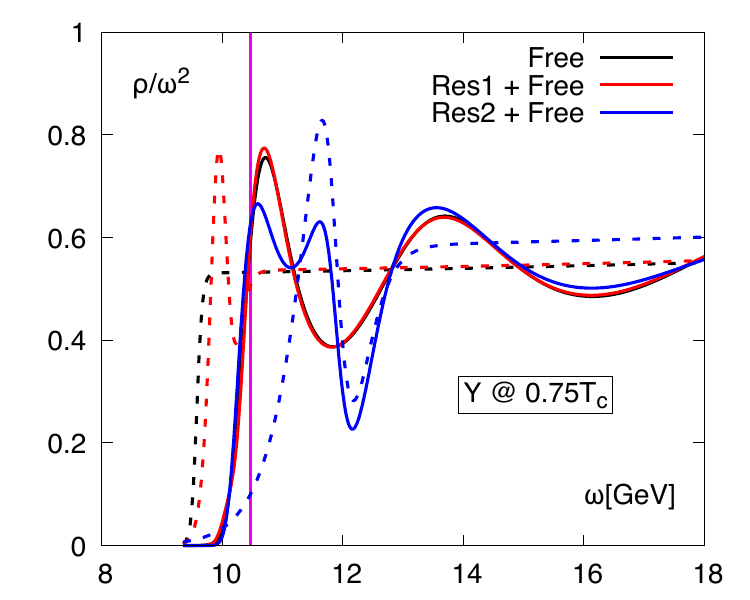}\includegraphics[width=0.33\textwidth,height=3.5cm]{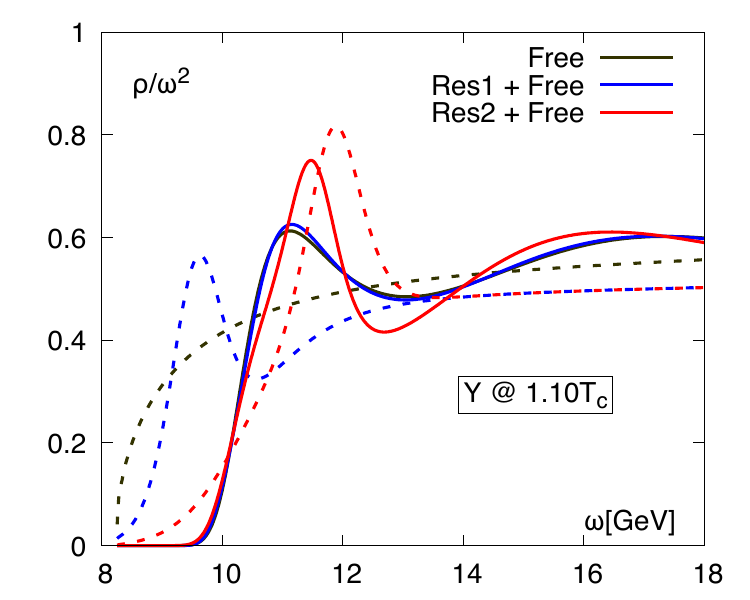}
     \includegraphics[width=0.33\textwidth,height=3.5cm]{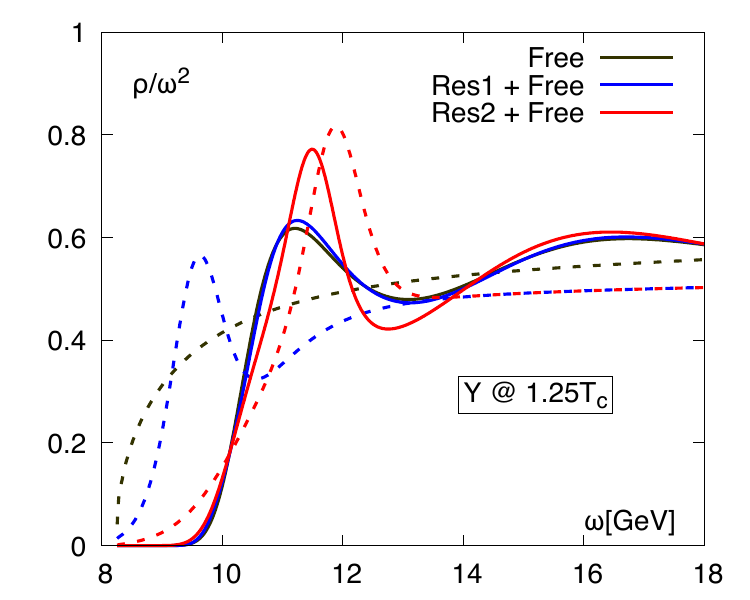}\includegraphics[width=0.33\textwidth,height=3.5cm]{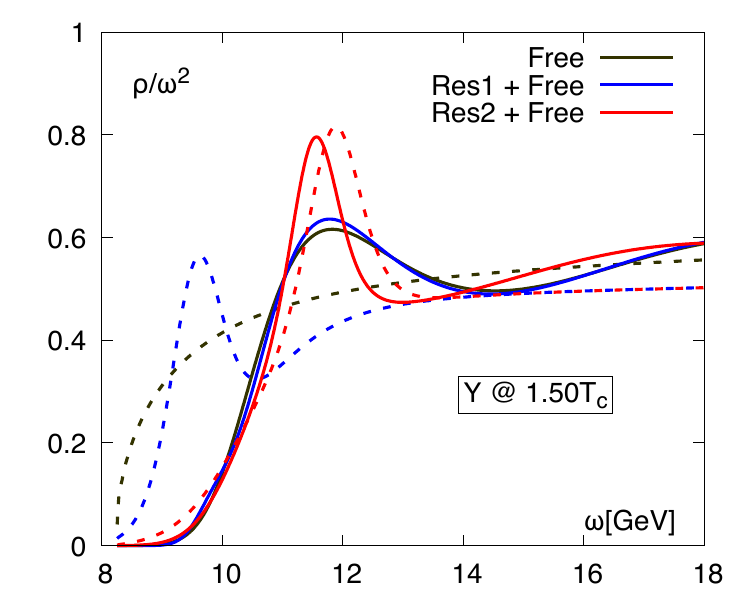}
  \end{center}\vspace{-0.5cm}
 \caption{The bottomonia spectral functions at different temperatures obtained by MEM with different default models. The vertical line indicates the screening mass of $\Upsilon$. \textit{Top left}: $0.75T_c$. \textit{Top right}: $1.10T_c$. \textit{Bottom left}: $1.25T_c$. \textit{Bottom right}: $1.50T_c$.}\label{bottom_dm}
 \end{wrapfigure}

As we have seen at 1.10 and 1.25$T_c$ a shift of the first resonance peak location has been observed. To see whether this is due to the reduction of number of data points at higher temperatures we compare the spectral functions extracted from the reconstructed correlation functions (denoted as ``$\rho$ from $G_{rec}$'' in Fig.\ref{charm_rec}) and the ones from the original correlation functions. We found that the shift still exists when the number of data points at the two temperates are the same. 

We also study the transport peak carefully at 1.10, 1.25 and 1.50$T_c$. At each temperature we use three different widths for the transport peak in the default model as input. We found that the width in the output spectral function is almost the same as the input. At each input width varying the height, however the output height firmly stays fixed. Increasing the input width the output height decreases which leads to a linear relation between $2\pi TD$ and $T/\eta$ (inverse width normalized by temperature) where $\eta$ is the drag force. We summarize the results in Fig.\ref{2piTD}. At 1.10, 1.25 and 1.50$T_c$ the obtained linear relations are $2\pi TD=0.789(11) T/\eta + 0.53(8)$, $2\pi TD=0.898(8) T/\eta + 0.43(2)$ and $2\pi TD=0.825(9) T/\eta + 0.54(7)$, respectively. To guide the eye we also show the Einstein relation at $M/T=2,6$.

Unlike the case in charmonia we extract bottomonia spectral functions from $G^{diff}(\tau/a)=G(\tau/a)-G(\tau/a+1)$ which will strongly suppress the contribution from the transport peak. Thus in our analysis for bottomonia we always use default models with threshold starting from a large frequency. The default models used for bottomonia are almost the same as those in the charmonia case except that no transport peak is used here and the resonance peak locations here will compare with the screening mass of $\Upsilon$. From the top left panel of Fig.\ref{bottom_dm} we see that at 0.75$T_c$, the first peaks in the output SPFs are independent of the default models and they are all close to the screening mass of $\Upsilon$. When temperature increases from $1.10T_c$ to $1.50T_c$, we found that for all cases the peak location of the low-lying peak has minor default model dependence and there is a small shift of first peak location for $T>T_c$ by around 0.52GeV.

 \setlength{\columnsep}{4pt}
\begin{wrapfigure}[11]{r}{0.66\textwidth}\vspace{-0.6cm}
\begin{center}
   \includegraphics[width=0.66\textwidth,height=2.8cm]{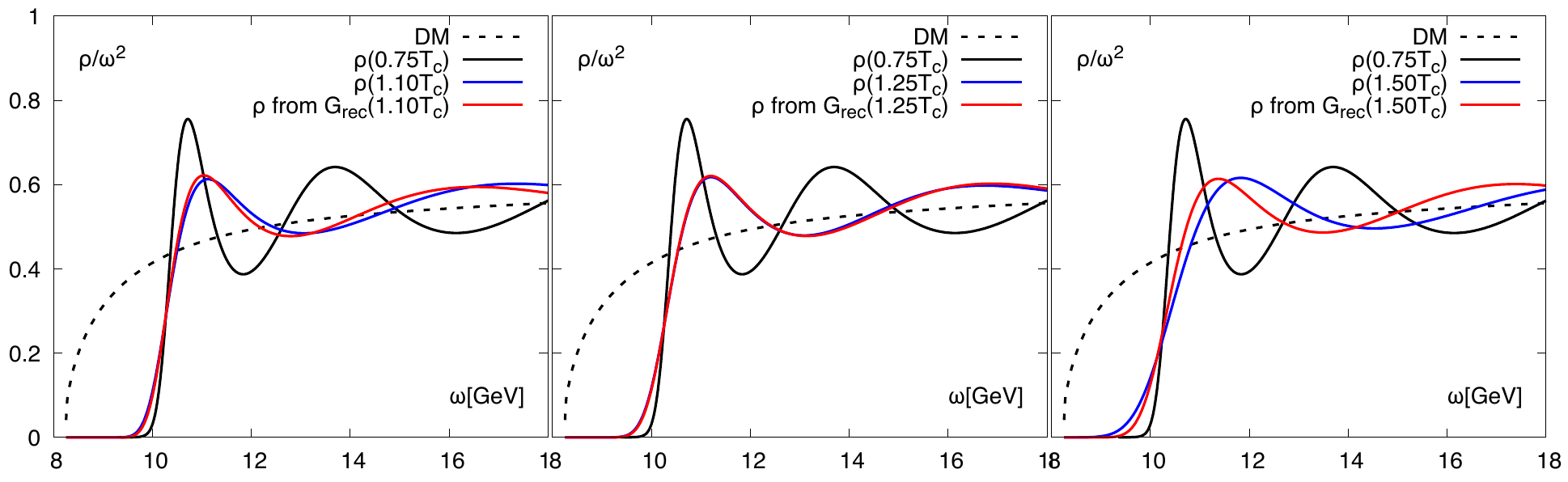}
\end{center}\vspace{-0.6cm}
   \caption{Comparison of bottomonia spectral functions in the vector channel obtained from the original correlation functions and the reconstructed correlation functions. \textit{Left}: $T=1.10T_c, T'=0.75T_c$. \textit{Middle}: $T=1.25T_c, T'=0.75T_c$. \textit{Right}: $T=1.50T_c, T'=0.75T_c$.}
   \label{bottom_rec}
 \end{wrapfigure}

The $N_{\tau}$ dependence of bottomonia are shown in Fig.\ref{bottom_rec}. We found that after getting rid of $N_{\tau}$ dependence the shift of first peak location is almost gone for all three temperatures. This indicates that $\Upsilon$ seems to stay unmodified up to 1.50$T_c$. Comparing the results obtained for charmonia and bottomonia, we can learn that $J/\psi$ suffers from more thermal modifications than $\Upsilon$. We have to mention that the results reported in this paper are based on the finest lattice data in~\cite{Ding:2017rty}. Our next step is to analyze the continuum-extrapolated data with physical charm and bottom masses.

This work is supported by the National Natural Science Foundation of China under grant numbers 11775096 and 11535012, the Deutsche Forschungsgemeinschaft (DFG) through the grant CRC-TR 211 ``Strong-interaction matter under extreme conditions'' and the U.S. Department of Energy, Office of Science, Office of Nuclear Physics, through Contract No. DE-SC001270 and Scientific Discovery through Advance Computing (ScIDAC) award ``Computing the Properties of Matter with Leadership Computing Resources''. The computations in this work were performed on the Aachen, Bielefeld, CCNU, Juelich and Paderborn machines.



\label{}






\bibliographystyle{elsarticle-num}
\bibliography{<your-bib-database>}



\end{document}